\documentclass{ws-procs9x6}
\def\>{\rangle}
\def\<{\langle}
\def\d{\hbox{d}}
\def\tr{\hbox{Tr}}
\begin{document}
\def\>{\rangle}
\def\<{\langle}
\def\d{\hbox{d}}
\def\tr{\hbox{Tr}}
\def\lb{\label}
\def\be{\begin{equation}}
\def\ee{\end{equation}}
\title{Can Black Hole Relax Unitarily?}
%{\footnote{Based on talks given at ``Black
%    Holes IV'', Honey Harbour, June 2003 and at BW2003, Vrnjacka Banja, August 2003.}}
\author{Sergey ~N. ~ Solodukhin}

\address{School of Engineering and Science, \\
International University Bremen, \\
P.O. Box 750561, Bremen 28759, Germany \\
E-mail: s.solodukhin@iu-bremen.de}

\maketitle

\abstracts
{We review the way the BTZ black hole relaxes back to 
thermal equilibrium after a small perturbation  
and how it is seen in the boundary (finite volume) CFT. 
The unitarity requires the relaxation  to be quasi-periodic.
It is preserved in the CFT  but  is  not obvious in the case of the semiclassical black hole 
the relaxation of which is driven by complex  quasi-normal modes. We discuss two
ways of modifying the semiclassical black hole geometry to maintain unitarity: 
the (fractal) brick wall and the worm-hole modification. In the latter case the entropy
comes out correctly as well.}

\section{Introduction} 
Any thermodynamical system initially in equilibrium at finite temperature 
and then perturbed tends to return to the equilibrium if the perturbation is not 
too big and does not last too long. The important parameter which
characterizes this process is the relaxation time $\tau$ \cite{Fetter}. In fact, the process
of relaxation back to the equilibrium is a particular and most easily tractable
example of more general phenomenon of the thermalization, when system
initially far from being thermal  gets thermalized to a state
characterized by certain  temperature. The way how the
thermalization goes for different systems is an important and still poorly
understood problem. Gravitational physics gives yet another example of system 
which behaves  thermally.  This system is black hole.
The black hole formation can be viewed as another example of the
thermalization: non-thermal collapsing body and the flat space-time geometry
in the beginning transform to (thermal) black hole state in the end of the
gravitational collapse. On the other hand, the already formed and  stayed in 
equilibrium black hole can be perturbed by exciting a pulse of matter field in
the exterior of black hole. 
The subsequent relaxation
is very well studied in the literature and is known to be characterized by
the so-called quasi-normal modes. These modes are eigen values of the radial 
Schr\"odinger type 
equation subject to certain boundary conditions.
These are dissipative boundary conditions saying that the perturbation should
leave the region through all possible boundaries. In general there are two
such boundaries: black hole horizon and spatial infinity. Formulated this way 
the boundary value problem is not self-adjoint so that the quasi-normal modes are
typically complex $\omega=\omega_R-i\omega_I$ with negative imaginary part.
In most cases there is discrete set of such modes parameterized by integer
number $n$. The imaginary value of the lowest $(n=1)$ quasi-normal mode
sets the relaxation time $\tau=1/\omega_I$. 

In fact what we have said in the beginning about the relaxation of any system
back to thermal equilibrium should be made more precise: the system should be in
infinite volume. In  finite volume and if the evolution of the system is
unitary, any perturbation   once created never leaves the system
so that the complete returning to the initial unperturbed state is not possible.
Thus information about the perturbation never disappears completely
and always can be restored.
The characteristic time  during which  the perturbation (as well as the whole
state of the system) is guaranteed 
to come back is set by the Poincar\'e recurrence time.
All this means that the characteristic frequencies which run the perturbation of 
unitary system in  finite volume should be real and discrete. Depending on these frequencies
the evolution of the perturbation is  quasi-periodic or even chaotic.
But it can never be dissipative. Thus, strictly speaking for the thermalization 
we need infinite volume. Of course, nothing is infinite in the real world.
The system still may  be considered as thermal during the interval of time which is
considerably less than the Poincar\'e recurrence time.

What this implies for the black holes? More specifically, for asymptotically AdS
black holes the state of thermal equilibrium of which is well defined
and can last infinitely long? Such a black hole can be viewed as system put in the
box with the size set by the AdS radius. So one would have to expect
this black hole to behave as any other system in the finite volume and 
in particular to show  the Poincar\'e recurrences (for the discussion of this in de
Sitter space see \cite{Susskind}). This however does not happen
in the semiclassical black hole: the complex quasi-normal modes are always
there. The presence of these modes is related to the very  existence  of the horizon.
Once there is black hole horizon there will always  be complex frequencies which govern
the time evolution of the perturbation. This problem is
a manifestation of the long-time debated issue of whether the black hole evolution is 
actually unitary (see \cite{Hawking} and \cite{Susskind}).

A refreshed look at the whole issue is offered by the AdS/CFT
correspondence (see review in \cite{Aharony:1999ti}). 
According to this correspondence the gravitational physics in the
bulk of asymptotically AdS space has a dual description in terms of a Conformal
Field Theory (CFT) living on the boundary. Thus, the black hole in the bulk
corresponds to a thermal CFT. The relaxation of the black hole than has a dual
description as relaxation of the CFT after a  perturbation driven by  certain
conformal operator has been applied to the system. The quasi-normal modes thus
set the time scale for the relaxation in the boundary CFT \cite{HH}. The effect of the
finite size however is rather delicate issue. It has been studied in paper
\cite{BSS} and is reviewed in section 3 of this note. The similar conclusions have been
made in paper \cite{Barbon}. The recent reviews on the issue of black hole
relaxation and unitarity are \cite{Sachs} 
and \cite{Barbon2}.

\section{Relaxation in black hole: quasi-normal modes} 
We consider (2+1)-dimensional BTZ black hole with metric given by
\be
ds^2=-\sinh^2y~ dt^2+dy^2+\cosh^2y~ d\phi^2,
\lb{1}
\ee
where for simplicity we consider non-rotating black hole and set the size of
the horizon $r_+=1$ and AdS radius $l=1$. The coordinate $\phi$ is periodic with period $L$ so
that the boundary has topology of cylinder and $L$ sets the finite size for
the boundary system.
A bulk perturbation $\Phi_{(m,s)}$ of mass $m$ and spin $s$
should satisfy the quasi-normal boundary condition, i.e. it should be in-going
at the horizon and have vanishing flux at the infinity. The latter condition
comes from the fact that in the asymptotically AdS space-times the effective
radial potential is growing at infinity so that there can  be no 
propagating modes as well as no leakage of the energy through the boundary.
The relevant radial equation takes the form of the hypergeometric equation
which is exactly solvable. The  quasi-normal modes in general fall into two
sets \cite{C-B,BSS2}
\begin{eqnarray}\label{2}
\omega&=&{2\pi\over L}m-4\pi i T_L(n+\bar h)\nonumber\\
\omega&=&-{2\pi\over L}m-4\pi i T_R(n+h)\ , m\in{\bf Z} \ , n\in{\bf N}
\end{eqnarray}
where the left- and right-temperatures $T_L=T_R=1/2\pi$ and $(h, \bar{h})$ 
have the meaning of the conformal weights of the dual operator ${\mathcal
  O}_{(h,\bar{h})}$ corresponding to the bulk perturbation
$\Phi_{(m,s)}$, with $h+\bar{h}=\Delta (m), \ h-\bar{h}=s$ and $\Delta(m)$ is
determined in terms of the mass $m$.

For comparison, in the case of global anti-de Sitter space the horizon and
respectively the quasi-normal modes are absent. But, instead, one can define
the normalizable modes which  form a discrete set of real frequencies\cite{Balasubramanian:1998sn}
\be
\omega=2\pi m/L+4\pi(n+h)/L~n\in {\bf N}~~
\lb{3}
\ee
where the size of the boundary is also set to be $L$ as in the black hole case.

\section{Relaxation in CFT$_2$}
The thermal state of the black hole in the bulk corresponds to  the thermal
state  on the CFT side. In fact, the boundary CFT factorizes on left- and
right-moving sectors with temperature $T_L$ and $T_R$ respectively. The bulk
perturbation corresponds to perturbing the thermal field theory state with
operator ${\mathcal O}_{(h,\bar{h})}$. The further evolution of the system is
described by the so-called Linear Response Theory (see \cite{Fetter}). 
According to this theory
one has to look at the time evolution of the perturbation itself.
More precisely, the relevant information is contained in the retarded
correlation function of the perturbation at the moments $t$ and $t=0$
(when the perturbation has been first applied). Since the perturbation is
considered to be small the main evolution is still governed by the unperturbed
Hamiltonian over the thermal state so that the correlation function is the
thermal function at temperature $T$.
Thus, the analysis boils down
to the study of the thermal 2-point function of certain 
conformal operators. Such a function should be double periodic: with period
$1/T$ in the direction of the Euclidean time and with period $L$ in the
direction of the compact coordinate $\phi$. This can be first calculated as a
2-point function on the Euclidean torus and then analytically continued to the
real time. 

\subsection{Universality}
In general the correlation function on torus can be  rather complicated since
its form is not fixed by the conformal symmetry. The conformal symmetry however
may help to deduce the universal form of the 2-point function in two special
cases: when size $L$ of the system is infinite (temperature $T$ is kept
finite)  and when inverse temperature is infinite (the size $L$ is finite). 
The universal form of the (real time) 2-point function  in the first case is
\be
\<{\mathcal{O}}(t,\phi ){\mathcal{O}}(0,0 )\> 
={(\pi T)^{2(h+\bar{h})} \over (\sinh \pi T (\phi-t
))^{2h}(\sinh \pi T (\phi+t ))^{2\bar{h}}}~~ \lb{31} \ee
which for large $t$ decays exponentially as 
$e^{-2\pi T(h+\bar{h}) t}$. The information about the perturbation is thus 
lost after characteristic time set by the inverse temperature. It is clear
that this happens because in  infinite volume the information may dissipate
to infinity. In the second case correlator
\be
\<{\mathcal{O}}(t,\phi ){\mathcal{O}} (0,0 )\> 
={(\pi/iL)^{2(h+\bar{h})} \over
(\sin \frac{\pi}{L}  (t+\phi ))^{2h}(\sin \frac{\pi}{L}  (t-\phi ))^{2\bar{h}}}~~
\lb{41}
\ee
has the oscillatory behavior. Notice that the oscillatory behavior in
the second case should have been expected since the system lives on the
circle. The perturbation once created at the moment $t=0$ at the point
$\phi=0$ travels around the circle with the speed of light and comes back to
the same point at $t=L$. Thus, the information about the perturbation is never
lost. The correlation function (\ref{41}) as a function of
time  represents a series of singular picks concentrated at $t=\pm \phi+nL$, $n\in
{\bf N}$. In fact, this behavior should be typical for any system with unitary
evolution in  finite volume.

It is interesting to see what happens in the intermediate regime when both $L$
and $1/T$ are kept finite. In this case the behavior of the correlation
functions is not universal, may depend on the (self)interaction in the system
and is known only in some cases. We consider two instructive examples: the
free fermion
field and the strongly coupled CFT which is dual to the gravity on AdS$_3$.

\subsection{Intermediate regime: Free fermions}
The two point
function of free fermions on the torus is known explicitly (e.g.
\cite{DiFrancesco:nk}). The real   time correlation function is
\be
\<\psi (w)\psi(0)\>_{\nu}={\theta_\nu (wT |i LT)\partial_{z}\theta_{1}(0|LT)\over
\theta_{\nu}(0|iLT)\theta_1 (wT |i LT)}~~,
\lb{5}
\ee
were $w=i(t+\phi)$ and $\nu$ characterizes the
boundary conditions for $\psi (w)$. For finite temperature boundary
conditions we have $\nu=3,4$. Using the properties of
$\theta$-functions, it is then easy to see that (\ref{5})
is invariant under shifts $w\rightarrow w+1/T$
and $w\rightarrow w+iL$.
It is then obvious that the resulting real time correlator
(\ref{5}) is a {\it periodic} function of $t$ with period $L$. Zeros of the
theta function $\theta_1(wT|iLT)$ are known \cite{DiFrancesco:nk}
to lie at $w=m/T +inL$, where $m,n$ are
arbitrary relative integers. Therefore, for real time $t$, the correlation
function (\ref{5}) is a sequence of singular peaks located at
$(t+\phi)=nL$. Using the standard representation \cite{DiFrancesco:nk}
of the $\theta$-functions, we also
find that in the limit $LT\rightarrow \infty$ the correlation function
(\ref{5}) approaches
the left-moving part of (\ref{3}) with $h=1/2$
that exponentially decays with time,
\be
\<\psi (w)\psi(0)\>_{3(4)}={\pi T\over 4\sinh{\pi T(t+\phi)}}[1\pm
2e^{-\pi LT}\cosh 2\pi T(t+\phi)+..]
\lb{61}
\ee
In the opposite limit, when $LT\rightarrow 0$,
it approaches the oscillating function (\ref{41}). 
A natural question is how the asymptotic behavior (\ref{61}) when size of the system is taken
to infinity can be consistent with the periodicity, $t\rightarrow t+L$,
of the correlation function (\ref{5}) at any finite $L$? 
In order to answer this question we have to observe that there are two
different time scales in the game. The first time scale is set by the inverse
temperature $\tau_1=1/T$ and is kept finite while the second time scale is associated
with the size of the system $\tau_2=1/L$. When $L$ is taken to infinity we
have that
$\tau_2>>\tau_1$. Now, when the time $t$ is of the order of $\tau_1$ but much
less than $\tau_2$ the asymptotic expansion (\ref{61}) takes place. The
corrections to the leading term are multiplied by the factor $e^{-\pi LT}$ and
are small. The 2-point function thus is exponentially decaying in this regime.
It seems that the system has almost lost   information about the initial
perturbation (at $t=0$). But it is not true: as time goes on and
approaches the second time scale $t\sim \tau_2$ the corrections to the
leading term in (\ref{61}) become important and the system starts to collect
its memory about the initial perturbation. The information is completely
recovered as $t=\tau_2$ and the time-periodicity is restored. 
This example is instructive. In particular, it
illustrates our point that there can be thermalization in the finite volume
for relatively small intervals of time, i.e. when $t<<\tau_2$.

\subsection{Strongly coupled CFT$_2$ dual to AdS$_3$} 
As an example of a strongly coupled
theory we consider the supersymmetric conformal field theory dual to string
theory on AdS$_3$. This theory describes the low energy excitations of
a large number of D1- and D5-branes \cite{Aharony:1999ti}. It can be interpreted as a gas of
strings that wind around a circle of length $L$ with target space $T^4$.
The individual strings can be simply- or multiply wound such that the total
winding number is ${\tt k}=\frac{c}{6}$, where $c>>1$ is the central charge.
The parameter $\tt k$ plays the role of N in the usual terminology of large N CFT.

According to the  prescription (see \cite{Aharony:1999ti}), each AdS space
which asymptotically approaches the given
two-dimensional manifold should contribute
to the calculation, and one thus has to sum over all such spaces.
In the case of interest, the two-manifold is a torus $(\tau, \phi )$,
where $1/T$ and $L$ are the respective periods.
There exist two obvious AdS spaces which approach the torus asymptotically.
The first is the BTZ black hole in AdS$_3$  and
the second is the so-called thermal AdS space,
corresponding to anti-de Sitter space filled with thermal radiation.
Both spaces can be represented (see \cite{Carlip:1994gc})
as a quotient of three dimensional
hyperbolic space $H^3$, with line element
\be
ds^2={l^2\over y^2}(dzd\bar{z}+dy^2)~~~y>0~~.
\lb{7}
\ee
In both cases, the boundary of
the three-dimensional space is a rectangular torus with
periods $L$ and $1/T$. We see that the two configurations
(thermal AdS and the BTZ black hole) are T-dual to
each other, and are obtained by the interchange of the coordinates
$\tau \leftrightarrow \phi$ and $L \leftrightarrow 1/T$ on the torus.
In fact there is a whole $SL(2,{\bf Z})$ family of spaces which are quotients
of the hyperbolic space.

In order to find correlation
function of the dual conformal operators, one has to solve the
respective bulk field equations subject to Dirichlet boundary
condition, substitute the solution into the action and
differentiate the action twice with respect to the boundary value
of the field. The boundary field thus plays the role of the source
for the dual operator ${\mathcal{O}}_{(h,\bar{h})}$.
This way one can obtain the boundary CFT correlation function for each member
of the family of asymptotically AdS spaces.
The total correlation function is then given by the sum over all $SL(2,{\bf
  Z})$ family with appropriate weight. For our purposes however it is
sufficient to consider the contribution of only two 
contributions \cite{Maldacena:2001kr} 
\be
\<{\mathcal{O}}(t,\phi) {\mathcal{O}} (0,0)\> \lb{8} \\
=
e^{-S_{\tt BTZ}}\<{\mathcal{O}} ~{\mathcal{O}}'\>_{\tt BTZ}+
e^{-S_{\tt AdS}}\<{\mathcal{O}} ~{\mathcal{O}}'\>_{\tt AdS}~~, 
\ee
where
$S_{\tt BTZ}=-{k \pi LT/2}$ and $S_{\tt AdS}=-{k  \pi /2LT}$
are Euclidean action of the
BTZ black hole and thermal AdS$_3$, respectively \cite{Maldacena:1998bw}.
On the Euclidean torus $\< \ \ \>_{\tt BTZ}$ and $\< \ \ \>_{\tt AdS}$ are T-dual to each other.
Their exact form can be computed explicitly \cite{Corrfunct}. 
For our purposes it is sufficient to note that the (real-time) 2-point function coming
from the BTZ part is exponentially decaying, $\< \ \ \>_{\tt BTZ}\sim e^{-2\pi
  hTt}$ 
even though it is a correlation function in a system of finite size
  $L$. On the other hand, the part coming from the thermal AdS is oscillating
 with period $L$ as it should be for a system at finite size. Thus, the total
2-point function (\ref{8}) has two contributions: one is exponentially
decaying and another is oscillating.  So that (\ref{8}) is {\it not} a
quasi-periodic function  of time $t$.
This conclusion does not seem to change if we include sum over $SL(2,{\bf Z})$ in
(\ref{8}).
There will always be contribution of the BTZ black hole that is exponentially decaying.
This can be formulated  also in terms of the poles in the momentum
representation of 2-point function (see \cite{BSS2} and
\cite{Danielsson:1999zt}). The poles of $\< \ \ \>_{\tt BTZ}$ 
are exactly the complex quasi-normal modes (\ref{2}) while that of 
 $\< \ \ \>_{\tt AdS}$  are the real normalizable modes (\ref{3}).

Depending on the value of
$LT$, one of the two terms in (\ref{8}) dominates \cite{Maldacena:1998bw}.
For high temperature ($LT$ is large)
the BTZ is dominating, while at low temperature  ($LT$ is small)
the thermal AdS is dominant. The transition between the two regimes
occurs at $1/T=L$. In terms of the gravitational physics, this corresponds to the
Hawking-Page phase transition \cite{Hawking:1982dh}.
This is a sharp transition for large $\tt k$, which is the case when the supergravity description is valid.
The Hawking-Page transition is thus a transition between oscillatory
relaxation at low temperature and exponential decay at high temperature.

\subsection{The puzzle and resolution}
Thus, the AdS/CFT correspondence predicts that the CFT dual to gravity on
AdS$_3$ is rather peculiar. Even though it is in finite volume the relaxation
in this theory is combination of oscillating and exponentially decaying functions.
This immediately raises a puzzle: how this behavior is consistent with the
general requirement for a unitary theory in finite volume to have only
quasi-periodic relaxation?
Resolution of this puzzle was suggested in \cite{BSS}. It was suggested that
additionally to the size $L$ there exists another scale in the game.  This
scale appears due to the fact that in the dual CFT at high temperature 
the typical configuration consists of multiply wound strings  which
effectively propagate in a much bigger volume, $L_{\tt eff}\sim {\tt k}L$.
The gravity/CFT duality however is valid in the limit of infinite $\tt k$
in which this second  scale becomes infinite. So that the exponential
relaxation corresponds to infinite effective size $L_{\tt eff}$
that is in complete agreement with the general arguments. At finite $\tt k$
the scale $L_{\tt eff}$ would be  finite and the correlation function 
is expected to be quasi-periodic with two periods $1/L$ and $1/L_{\tt eff}$.
The transition of this quasi-periodic function to combination of exponentially
decaying and oscillating functions when $L_{\tt eff}$ is infinite then should be  similar to what we
have observed in the case of free fermions when $L$ was taken to infinity.

\section{ Black hole unitarity: finite $\tt k$}
That relaxation of black hole is characterized by a set of complex frequencies
(quasi-normal modes) is mathematically precise formulation of the lack of
unitarity in the semiclassical description of black holes.
The unitarity problem was suggested to be resolved within the AdS/CFT
correspondence \cite{Maldacena:2001kr}. Indeed, the theory on the boundary is unitary and there should
be a way of reformulating the processes happening in the bulk of black hole space-time
on the intrinsically  unitary language of the boundary CFT. The analysis of the relaxation
is helpful in understanding how this reformulation should work.
Before making comments on that let us note that the loss of information
in semiclassical black hole is indeed visible on the CFT side.
It is encoded in that exponentially decaying contribution to the 2-point
correlation function. For the CFT itself this however is not a problem.
As we discussed above the finite size unitarity  is restored at finite value
of $\tt k$. This however goes beyond the limits where the gravity/CFT duality
is formulated. Assuming that the duality can be extended to finite
$\tt k$ an important question arises: {\it What would be the gravity counter-part
  of the duality at finite $\tt k$?} Obviously, it can not be a semiclassical
black hole. The black hole horizon should be somehow removed so that the
complex quasi-normal modes (at infinite $\tt k$) would be replaced by
real (normal) modes when $\tt k$ is finite. Below we consider two
possibilities of how it may happen.

\subsection{Fractal brick wall}
It was suggested in  \cite{Barbon} that the quantum modification of the black
hole geometry needed for the restoring the Poincar\'e recurrences can be modeled 
by the brick wall. Here we elaborate on this interesting idea.
The brick wall is introduced by placing a boundary at small
distance $\epsilon$ from the horizon and  cutting off a part of the
space-time lying inside the boundary. The effect of the boundary on the
quantum fields is  implemented by imposing there the Dirichlet boundary
condition. Originally, the brick wall was introduced by 't Hooft \cite{tHooft} for
regularizing the entropy of the thermal atmosphere  out-side black hole horizon.
With this regularization the quantum entropy $S_{q}\sim {A\over \epsilon^{d-2}}$
correctly reproduces the proportionality of the black hole entropy to the
horizon area $A\sim r_+^{d-2}$. Assuming that $\epsilon$ is taken to be of the
order of the Planck
length, so that Newton's constant is $G\sim \epsilon^{d-2}$, one can argue
that the black hole entropy is correctly reproduced in this approach.
Later on it was however realized that the brick wall divergence is
actually a UV divergence. One can introduce a set of the 
Pauli-Villars fields with masses set by parameter  $\mu$  which plays the role of the UV regulator. 
Taking into account the contribution of the regulator fields
in the entropy of the quantum atmosphere the brick wall 
can be removed \cite{Myers}. The entropy then is proportional to certain power of the UV
regulator, $S_q\sim A\mu^{d-2}$. 

In our story of  black hole relaxation the brick wall indeed gives the wanted
effect: once the brick wall has been introduced the quasi-normal modes
disappear completely
and are replaced by a set of the real (normal) modes. This happens because the
effective infinite size region near horizon is now removed and the whole
space  is the finite size region between the brick wall and the boundary at
spatial infinity. In such a system we expect periodicity with the period set by the
brick wall parameter $\epsilon$ as $t_{\tt bw}\sim 1/T\ln (1/\epsilon)$. This
periodicity shows up in the boundary CFT correlation functions rather
naturally. Indeed, these correlation functions are constructed from the bulk
Green's function which describes propagation of the perturbation between two
points on the boundary through the bulk. In the present case the perturbation 
from a point $\phi$ on the boundary goes along null-geodesic
through the bulk, reflects at the brick wall and
returns to the same point $\phi$
on the boundary. The time which the perturbation travels gives the periodicity
for the boundary theory and it equals $t_{\tt bw}$. Matching $t_{\tt bw}$ and
$1/L_{\tt eff}$ gives the relation between brick wall regulator $\epsilon$ and 
parameter $\tt k$ of the large N boundary CFT. 

This probably should be  enough for 
the explaining and reproducing the second time scale of the boundary CFT from
the gravity side. The time $t_{\tt bw}$ is however much smaller than the
Poincar\'e recurrence time which is expected to be of the order, $t_{\tt P}\sim e^{A\over
  \epsilon}$. So how to get this time scale in the model with the brick wall?
We notice that the brick wall should not be ideally spherical. The possible complexity of
the shape is not restricted. It may even be fractal. In order to serve as a
regulator for the quantum entropy calculation brick wall should just stay at mean
distance $\epsilon$ from the horizon but its shape can be arbitrary.
For the recurrence time the shape is however crucial. In the absence of the
spherical symmetry the perturbation emitted from the point $\phi$ on the boundary 
(which is still a circle) at spatial infinity goes along null-geodesic through the
bulk, reflects from the brick wall, goes back and arrives at completely
different point $\phi'$ on the boundary at spatial
infinity. Only after a number of  back and forth goings between two boundaries
the perturbation can manage to arrive on the boundary at the same point where
it was initially emitted.  This number can be very large and it sets the
periodicity for the boundary theory. The emerging geometric picture is standard
set up for the system having  classical chaos. Indeed, generic deviations from
the spherical symmetry of one of the boundaries leads to chaotic behavior of
the geodesics. 
This means that the 2-point functions on the boundary would
generically have chaotic time evolution. The optical volume $V$ between two boundaries
seems to be the right quantity to measure the size of the phase space of the
chaotic geodesics. Since $S_q\sim V$ the recurrence time $t_{\tt P}\sim e^V$
gives the right estimate for the Poincar\'e time.
In this picture the information sent to black hole eventually comes
back. The characteristic time during which it should happen is set by the
Poincar\'e recurrence time $t_{\tt P}$.

The classical chaos of the geodesics  manifests  in the (normal) frequencies.
The latter are the eigen values of the Laplace-type operator  considered on the 
classical geometry. As we know from the relation between classical and quantum
chaos the chaos of the geodesics in the classical system manifests in that the
eigen values of the quantum problem are randomly 
distributed. Thus, the normal frequencies will be random numbers. This again
means that the 2-point function on the boundary (we  expect that the
normal modes are still poles in the momentum representation of the correlation
function) is chaotic function of time.

The irregularity of the shape of the brick wall may actually  be physically
meaningful. It can model the
fluctuating quantum horizon. It may also be a way of representing the
so-called stretched horizon (see \cite{Kabat}).

\subsection{Worm-hole modification: BTZ$_{\tt k}$}
The horizon can be removed in a smooth way by modifying the black hole
geometry and making it  look like a worm-hole. As an example we present here 
a modification of the BTZ metric (\ref{1}),
\be
ds^2=-(\sinh^2y+{1\over {\tt k}^2})~ dt^2+dy^2+\cosh^2y~ d\phi^2,
\lb{9}
\ee
which we call BTZ$_{\tt k}$. The horizon which used to stay at $y=0$
disappears in metric (\ref{9}) if $\tt k$ is finite. The whole geometry now is that of worm-hole 
with the second asymptotic region at $y<0$. 
The two asymptotic regions separated by horizon in classical BTZ metric can
now talk to each other leaking the information through the narrow throat.
The metric (\ref{9}) is still asymptotically AdS although it is no more a constant
curvature space-time. The Ricci scalar
\be
R=-{2\over ({\tt k}^2 \sinh^2y+1)^2}[({\tt k}^2+1)+3{\tt k}^4\sinh^4y+5{\tt
  k}^2\sinh^2y ]
\lb{10}
\ee
approaches value $-6$ at infinite $y$ and $-2({\tt k}^2+1)$ at $y=0$ where the
horizon used to stay.
The normal frequencies in the space-time with metric (\ref{9}) 
are real and are determined by the normalizability and the Dirichlet boundary
condition at both spatial infinities.  Since
the space-time (\ref{9}) is asymptotically AdS one can use the rules of the AdS/CFT
duality and calculate the boundary correlation function. Technically it is
more difficult than in the standard BTZ case since (\ref{9}) is not maximally
symmetric space. But the result should be a periodic in time function with the
period set by parameter $\tt k$. It would be interesting to do this
calculation and  see if this correlation function makes sense from the point
of view of the expected behavior of the boundary CFT at finite $\tt k$.
One can calculate the entropy of the thermal atmosphere in the metric
(\ref{9}). It is now finite with no need for introducing the brick wall. 
The entropy then behaves as $S_q\sim {\tt k}A$ that is the right answer  for the 
Bekenstein-Hawking entropy of BTZ black hole. Thus, the modification (\ref{9}) gives us the
right entropy and solves the unitarity problem. 

\medskip

%\section*{Acknowledgments}
\noindent{\bf {Acknowledgments}} I would like to thank D. Birmingham and I. Sachs for enjoyable collaboration 
and many useful discussions.
I also thank G. Arutyunov, J. Barbon, A. Morozov and N. Kaloper for
important discussions.

\end{document}